\documentclass[12pt,preprint]{aastex}
\usepackage{emulateapj5}
\begin{document}

\newcommand{\kms}{\mbox{km~s$^{-1}$}}
\newcommand{\s}{\mbox{$''$}}
\newcommand{\mloss}{\mbox{$\dot{M}$}}
\newcommand{\my}{\mbox{$M_{\odot}$~yr$^{-1}$}}
\newcommand{\ls}{\mbox{$L_{\odot}$}}
\newcommand{\ms}{\mbox{$M_{\odot}$}}
\newcommand\mdot{$\dot{M}  $}
\def\arcdeg{\hbox{$^\circ$}}
\def\12co{$^{12}$CO}
\def\mloss{\mbox{$\dot{M}$}}
\def\my{M$_\odot$ yr$^{-1}$}
\def\etal{{\it et al.~\/}}
\def\mpc{$\ {h^{-1}\rm Mpc}$}
\def\um{$\,\mu$m}
\def\f{$f_{\nu}(\nu_{60})$}
\def\arcsec{\ifmmode {^{\prime\prime}}\else $^{\prime\prime}$\fi}
\def\arcmin{\ifmmode {^{\prime}}\else $^{\prime}$\fi}
\def\iras{{\sl IRAS\/}}
\def\simlt{\lower.5ex\hbox{\ltsima}}
\def\simgt{\lower.5ex\hbox{\gtsima}}
\def\ltsima{$\; \buildrel < \over \sim \;$}
\def\gtsima{$\; \buildrel > \over \sim \;$}
\def\ne2{[Ne\,II]}

\title{X-RAY EMISSION FROM THE PRE-PLANETARY NEBULA HENIZE 3-1475}

\author{Raghvendra Sahai\altaffilmark{1}, Joel H. Kastner\altaffilmark{2},  
Adam Frank\altaffilmark{3}, Mark Morris\altaffilmark{4}, Eric G. Blackman\altaffilmark{3}
}

\altaffiltext{1}{Jet Propulsion Laboratory, MS\,183-900, Caltech,
    Pasadena, CA 91109}

\altaffiltext{2}{Chester F. Carlson Center for Imaging Science, Rochester Institute of 
Technology, Rochester, NY 14623}

\altaffiltext{3}{Dept. of Physics \& Astronomy, University of Rochester, Rochester, NY 14627}

\altaffiltext{4}{Division of Astronomy \& Astrophysics, UCLA, Los Angeles, CA 90095}

\email{raghvendra.sahai@jpl.nasa.gov}

\begin{abstract}
We report the first detection of 
X-ray emission in a pre-planetary nebula, Hen\,3-1475. Pre-planetary nebulae 
are rare objects in the short transition stage between the Asymptotic Giant Branch 
and planetary nebula evolutionary phases, and Hen\,3-1475, 
characterised by a remarkable S-shaped chain of optical knots, is one of the 
most noteworthy members of this class. 
Observations with the Advanced CCD Imaging Spectrometer (ACIS) onboard the 
Chandra X-Ray observatory show the presence of  
compact emission coincident with the brightest optical knot in this bipolar object, 
which is displaced from the central star by 2\farcs7 along the polar axis.
Model fits to the X-ray spectrum indicate an X-ray temperature and luminosity, 
respectively, of $(4.3-5.7)\times10^6$\,K 
and $(4\pm1.4)\times10^{31}$ (D/5\,kpc)$^2$ erg\,s$^{-1}$, respectively. 
Our 3$\sigma$ upper limit on the luminosity of compact X-ray emission from the 
central star in Hen\,3-1475 is $\sim 5 \times 10^{31}$ 
(D/5\,kpc)$^2$ erg\,s$^{-1}$.
The detection of X-rays in Hen\,3-1475 is consistent with models in which 
fast collimated post-AGB outflows are crucial to the shaping of planetary nebulae; 
we discuss such models in the context of our observations.

\end{abstract}


\section{Introduction}
Pre-Planetary nebulae (PPNs)\footnote{we prefer this phrase to ``proto-planetary" nebula, 
which although in more common use for these objects, is also potentially misleading because the 
same term is used by the young stellar object community} -- objects in transition between the 
Asymptotic Giant Branch and planetary nebula (PN) phases -- hold the key to 
some of the most vexing problems in our understanding of these very late 
stages of stellar evolution for intermediate mass ($\sim 1-8$\ms) stars. 
Observationally, PNs show a dazzling variety 
of morphologies \cite[e.g.,][]{schwarz92,st98}, whereas AGB stars are 
surrounded by roughly spherical gas-dust envelopes due to dense, 
slow stellar winds ejected with rates up to 10$^{-4}$\my \cite[e.g.][]{neri98}. 
In the short transition period between the AGB and PN phases, 
these objects turn on very fast winds (with speeds of a  
few$\times$100 to $\sim$2000\,\kms) which interact violently with the 
surrounding envelopes, 
and drastically modify their spatio-kinematic structures. The nature and 
origin of such outflows remains unknown, and their interaction with the 
ambient circumstellar material and the consequent changes in the 
kinematics and morphology of the circumstellar material are very poorly 
understood.

X-ray (0.1-10 keV) observations can probe the direct 
signature of the interaction of the fast and slow winds, i.e. the very 
hot (10$^{7-8}$K) gas which should be produced in the shocked region. 
As a result, many searches for X-ray emission from PNs have been conducted 
beginning with the Einstein observatory 
(mid-1980s), followed by EXOSAT and ROSAT, and continue now with CXO and 
XMM-Newton. Prior to ROSAT, most X-rays in PNs were interpreted as 
coming from the hot (1-2$\times10^5$K) central star; ROSAT data led to the 
identification of {\it diffuse} emission from 7 PNs but several of these 
were erroneous (Kreysing et al. 1992, Guerrero, Chu \& Gruendl 2000). 
With CXO and XMM, extended X-ray emission 
from PNs has been unambiguously imaged, and results have been reported for 
5 objects (BD+30\arcdeg3639, NGC7027, Mz\,3: Kastner et al. 
2000,2001,2003; NGC6543: Chu et al. 2001; NGC7009: Guerrero, Gruendl \& Chu 2002). 
These studies clearly show the 
presence of extended bubbles of hot ($\sim$10$^6$K) gas, in qualitative 
agreement with our expectations from interacting-winds models. 

X-ray observations of PPNs provide us with a unique opportunity to 
study magnetic fields, which 
may play an important role in shaping planetary nebulae (e.g. Balick \& Frank 2002), 
indirectly (acting as the collimation agent of the fast outflows: 
Garcia-Segura 1997, Gardiner \& Frank 2001) or directly 
(magneto-centrifugal launching of disk winds: Blackman, Frank and Welch 
2000). Unlike most PNs, the central stars of PPNs are 
too cold to provide blackbody X-ray emission, thus detection of a compact 
central source of X-rays in a PPN could provide direct evidence for the presence of a strong 
stellar magnetic field. E.g., 
by analogy with the solar dynamo, new AGB-stellar-dynamo modelling 
\citep{blackman01} predicts a strong field (of few$\times10^4$ G) 
near the base of the star's convection zone with a 0.2 yr decay half-cycle 
(11 yrs for the Sun); dissipation of this field should result in a 
non-thermal X-ray luminosity of $\sim 10^{32}$ erg\,s$^{-1}$ from the central 
stars in very young PPNs. 

However, no systematic search for X-ray emission from PPNs has been carried 
out to-date, because until very recently, it was believed that the wind-wind 
interaction responsible for shaping PNs occurred during the PN phase 
(through the interaction of a fast radiatively-driven spherical wind from 
the central hot white dwarf with an equatorially-dense wind from the AGB 
phase). In this {\it Letter}, we report our discovery of 
X-ray emission from the well-studied PPN Hen\,3-1475. This remarkable S-shaped object 
was selected as the prime target in our search for X-ray emission from PPNs 
because it shows the presence of an ultra-fast 
outflow ($\sim$2300\,\kms) \citep{ss01}, and  
bright optical knots of emission from shocked gas \citep{bh01,riera03}. 


\section{Observations \& Results}

We observed Hen\,3-1475 with the Chandra X-Ray Observatory (CXO) 
for 49.7 ks on 2002 July 15, using the back-illuminated CCD S3 of the
Advanced CCD Imaging Spectrometer (ACIS). The CXO/ACIS combination is sensitive
over the energy range 0.3-10 keV. ACIS has a pixel size of
0.49$''$, very similar to the width of the point spread function (PSF)
of the Chandra mirrors. The data, consisting of individual CCD
X-ray and particle events, were subject to standard processing by Chandra
X-ray Center pipeline software
(CIAO\footnote{http://cxc.harvard.edu/ciao/}, version 2.0), which
determines the distribution of photon-generated charge within a 3$\times$3
CCD pixel box centered on the event position, flags events likely due
to particles, and computes the celestial positions and nominal energies of
incident X-rays. In order to optimize image resolution, 
the event charge distributions were used, along with the telescope pointing history 
and nominal photon energies, to calculate a subpixel position for each X-ray 
\citep{li03}.

\begin{figure}[htb]
\vskip -0.3in
\hskip -0.2in
\includegraphics[scale=0.8]{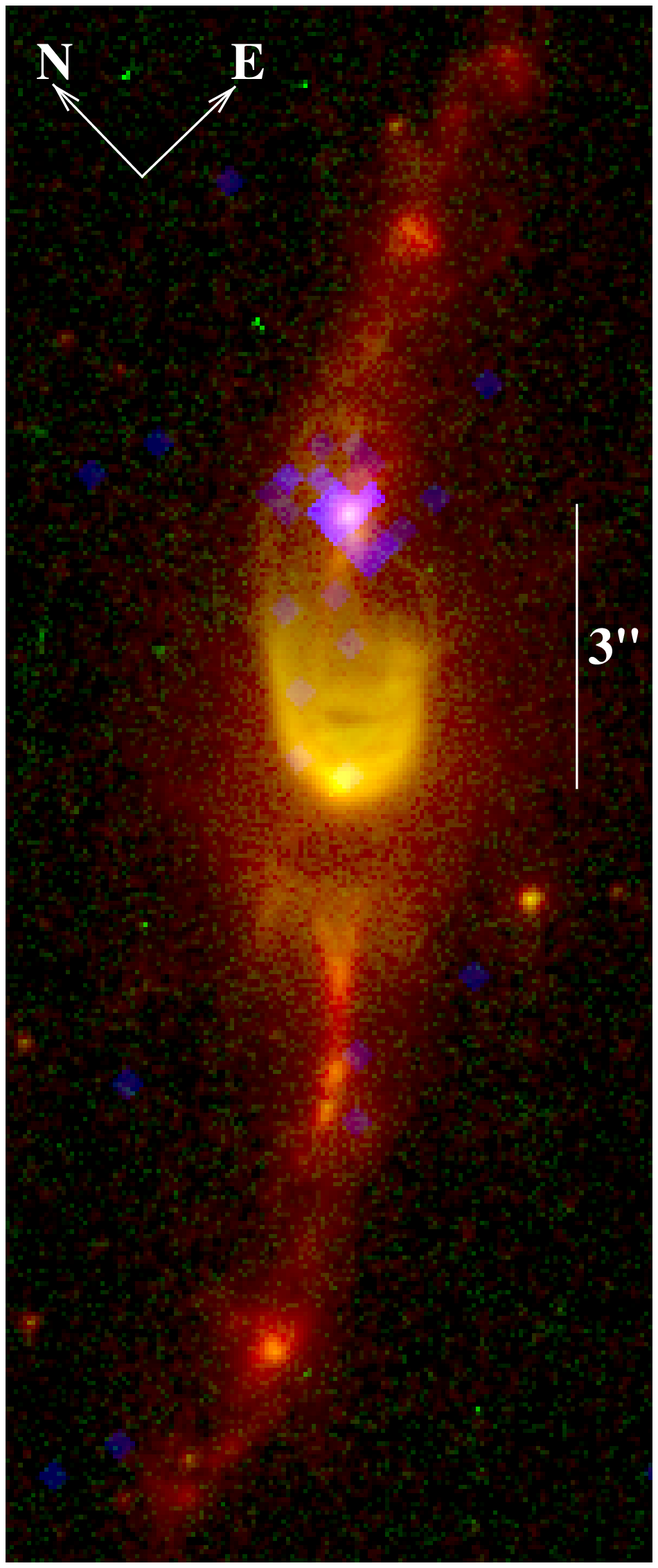}
\caption{CXO/ACIS-S3 X-ray image (blue, linear stretch) overlaid on a composite HST image 
of Hen\,3-1475 taken through the F507N ([OIII], green, log stretch) and F658N ([NII], red, log stretch) 
emission-line filters
}
\label{f1_x_oiii_nii}
\end{figure}

The CXO/ACIS-S3 image of Hen\,3-1475 (Fig. \ref{f1_x_oiii_nii}) shows the presence of 
compact emission coincident with the brightest optical knot in this object 
(NW1a in Fig. 1, \citep{riera03}), 
as seen in a Hubble Space Telescope (HST) image taken with the Wide Field \& Planetary Camera 2 (WFPC2) 
in the [OIII] (F502N) filter. The relative registration of the optical and X-ray images has 
been established as follows: (1) the USNO-B.1 catalog was used to determine the optical position 
of the central star in Hen\,3-1475 ($\alpha=17^h45^m14^s.186, \delta=-17\arcdeg56\arcmin46\farcs23$, 
J2000), and (2) the absolute CXO astrometry was used for the X-ray data, which locates 
the central peak of the X-ray emission at 
$\alpha=17^h45^m14^s.05, \delta=-17\arcdeg56\arcmin44\farcs81$, i.e. 
offset by 2\farcs0\,W, 1\farcs4\,N from the 
central star. The peak of the optical (NW1a) knot is 
located 1\farcs95\,W, 1\farcs9\,N, of the central star. Taking into account (a) the size of 
the optical knot (about 0\farcs7 along the nebular axis), (b) the sizes of the CXO PSF and the ACIS-S3 pixel 
size (about 0\farcs5), and (c) the uncertainty in the 
CXO absolute astrometry ($\sim$0\farcs6), we believe that the X-ray emission 
and the optical knot are co-spatial, and thus physically associated.
The probability of chance association of a background X-ray source with the knot is 
very small, considering that the optical knot covers only $\sim10^{-5}$ of 
a region of $\sim$2 arcmin$^2$ around the source in which there are 
only a few ($<$5) X-ray sources.

\begin{figure}[htb]
\vskip -0.3in
\resizebox{1.0\textwidth}{!}{\includegraphics{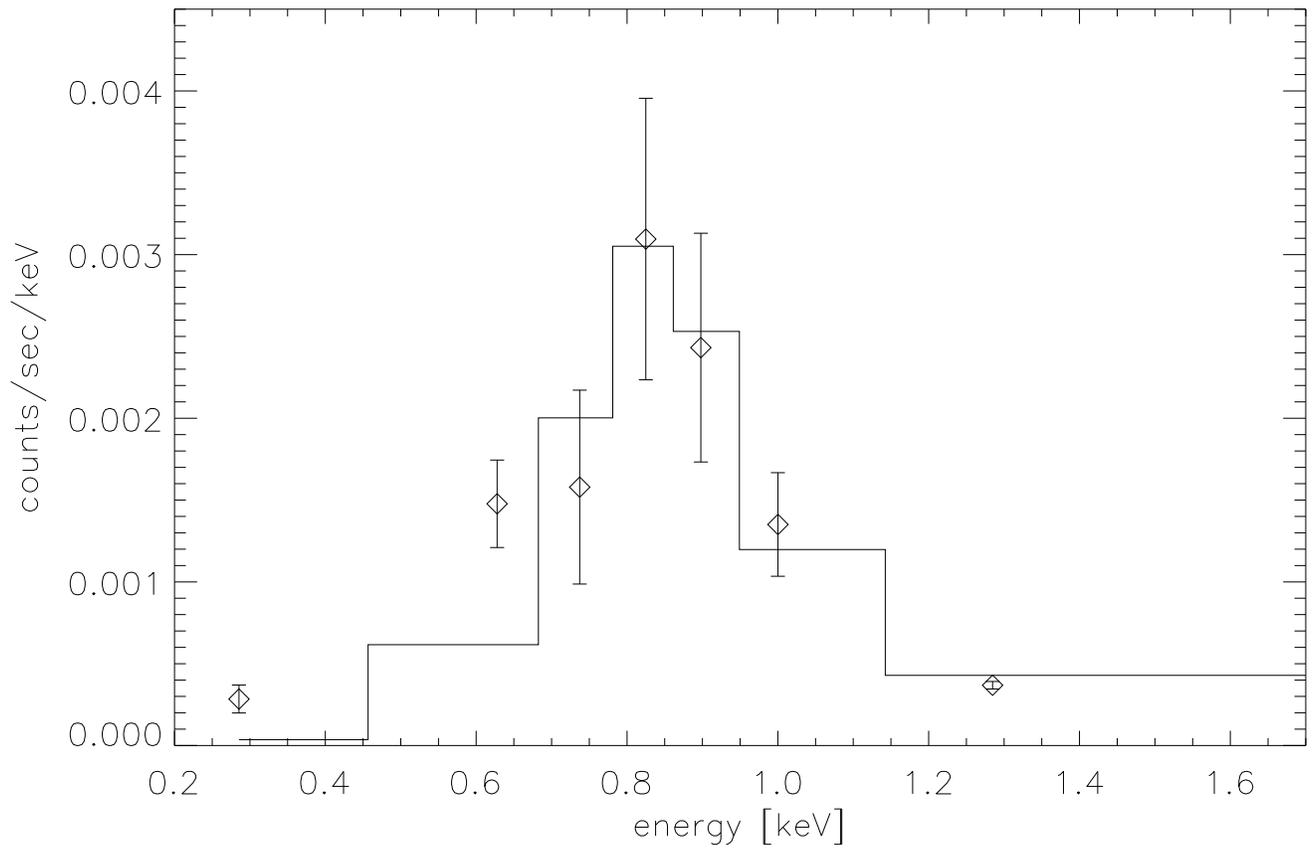}}
\caption{ACIS-S3 (background-subtracted) spectrum of Hen\,3-1475, together with a best-fit VMEKAL 
plasma model with $T_x=5\times10^6$\,K
}
\label{spec}
\end{figure}

The spectrum of the X-ray emitting knot 
is fairly soft, with almost all the flux confined to the 0.4--1.6 
keV energy band (a total of 63$\pm$8 counts, corresponding to a background-subtracted count rate 
of 1.5$\times10^{-3}$ cps) (Fig. \ref{spec}). The spectrum was extracted using CIAO tools (ver. 2.3), 
a 5 pixel (2\farcs5) radius circular source region centered on the X-ray knot 
and a somewhat larger circular background region centered
$\sim20''$  west of the source. The spectrum peaks around 0.85 keV, 
similar to the ACIS-S spectrum of Mz\,3 \citep{joel03mz3}. In contrast, two background 
sources in the vicinity of Hen\,3-1475 
have significantly harder spectra. 

We have fitted the X-ray spectrum with a model consisting of VMEKAL thermal emission plus
``standard ISM" intervening absorption \citep{morrison83}; a value of $A_V$=2 is 
assumed for the latter \citep{riera95}. Assuming 
solar abundances, the values of the plasma temperature 
kT and the source flux were varied; the best-fit, assuming a 20\% uncertainty 
in $A_V$, gives a temperature, $T_x=(4.3-5.7)\times10^6$\,K, 
and an ``unabsorbed" X-ray flux, $F_x=(0.87-1.8)\times10^{-14}$\,erg cm$^{-2}$ s$^{-1}$, where 
the lower (upper) limit for $F_x$ corresponds to the upper (lower) limit for $T_x$.

Although the above model provides a reasonable fit to the observed 
spectrum, the latter shows an excess (relative to the model) around 0.65 keV. Since 
oxygen (O VIII) can contribute to this feature via line emission at 
19\AA, we have tried models in which the O abundance is allowed to 
vary freely. We find that there is a hint of
O enhancement in the X-ray-emitting gas, but the signal-to-noise ratio is insufficient 
to determine abundances. Unless, e.g., Fe is severely depleted in Hen\,3-1475(as 
may be the case for BD+30\arcdeg3639; Maness et al. 2003), the 
best-fit value of $T_x$ obtained from the
solar abundance model is likely to be reasonably accurate. The X-ray 
flux ($F_x$) determination is fairly robust even if heavy metals are
severely depleted.

Our derived $F_x$ gives an intrinsic X-ray luminosity for 
Hen\,3-1475 of $L_x=(4\pm1.4)\times10^{31}$ (D/5\,kpc)$^2$ erg\,s$^{-1}$.
Using our model volume emission measure, EM$\sim 10^{54}$ cm$^{-3}$, and the electron density 
for the NW knot ($\sim$ 3000 cm$^{-3}$) from \cite{riera95}, we find that the 
X-ray emitting volume is  
$V_x \sim 10^{47}$ cm$^{3}$, which is a small fraction 
of the optically emitting volume ($V_{opt} \sim 
3\times10^{48}$ cm$^{3}$), as derived from the knot size ($\sim$0\farcs2) in the [OIII] image. 

It is not surprising that X-ray emission is detected only from 
the NW1a knot, since this feature 
is also the brightest and highest-excitation optical jet feature, by far, as seen in the 
[OIII]$\lambda$5007 image (Fig. \ref{f1_x_oiii_nii}). For example, the integrated [OIII] flux 
and the [OIII]/[NII] emission-line ratio of 
knot NW1a, are larger by factors 3.5 and $\gtrsim$2, respectively, 
than the corresponding values for its point-symmetric counterpart (knot SE1a). 
The 1.4 mag of differential extinction 
between the knots ($A_V$=3.4 for SE1 knots, \cite{riera95}) reduces the X-ray flux of the SE1a knot 
by a factor 6.5 over the NW1a knot, insufficient to make it undetectable in our data.
Thus, the difference in the X-ray emission 
(and optical emission) between knot NW1a and its SE counterpart probably results from 
a combination of differences in extinction and excitation. 


We do not detect a compact X-ray source towards the central star in Hen\,3-1475; a conservative 
upper limit (3$\sigma$) on the ``unabsorbed" X-ray flux and intrinsic X-ray luminosity, 
respectively, of such a source is 
$F_c \sim 1.7\times10^{-14}$\,erg cm$^{-2}$ s$^{-1}$ and $L_c \sim 5 \times 10^{31}$ 
(D/5\,kpc)$^2$ erg\,s$^{-1}$, 
using a temperature similar to that of the nebular source, and 
$A_V=3.4$, the value used \cite{riera95} for the central source 
in Hen\,3-1475. Since the extinction towards the center could be higher, and the 
distance to Hen\,3-1475 is rather uncertain, our upper limit is not inconsistent with the 
value predicted by \cite{blackman01} ($\sim 10^{32}$ erg\,s$^{-1}$) based on their 
solar-dynamo analog for producing magnetic fields in the central stars of PPN.

\section{Discussion}
Recent HST studies of the 
morphology of young PNs and PPNs strongly support a model in which fast 
collimated outflows begin to modify the circumstellar environment during 
the PPN phase and/or even earlier, during the very late AGB phase (Sahai 
\& Trauger 1998). Our discovery of nebular 0.4-1.6 keV X-ray emission 
from Hen\,3-1475 dramatically confirms the operation of the resulting violent 
post-AGB wind/ AGB wind interaction process. 
The simultaneous presence of both X-ray and optical emission in the NW1a knot 
supports the idea that the knots are strongly-cooling 
regions of shocked gas with very strong temperature gradients.
The S-shaped distribution of the fast-moving knots in this object 
indicates that the fast post-AGB wind consists of an episodic 
jet-like outflow whose axis has precessed with time (precession angle $<$10\arcdeg~and 
a period $\sim$1500 years, \cite{riera95}). This scenario is supported 
by the theoretical result that a jet with a quasi-periodic variability and slow 
precession fragments into clumps which behave like ``interstellar bullets" \citep{rb93}.

Alternatively, the jet may intrinsically consist of a 
series of hypersonic bullets ejected by processes close to the surface of
the star.  \cite{matt04} have shown how the development of toroidal
fields in a recently exposed PPN core could lead to an impulsive
deposition of flow energy into a collimated system. Whatever their
origin, recent work by \cite{polud04} has shown that
hypersonic bullets are capable of recovering key aspects of many
PPN flows. The temperatures deduced from the X-ray measurements
could be achieved with shock velocities at the surface of the
bullet of $ v_s \approx \sqrt{(16 k T_x /3 m_h)}$ = 470\,\kms~.  
Note that although this speed is lower than that derived for the 
knots from observations \citep{bh01,riera03}, the relevant speed for comparison 
is that of the shock transmitted into the high density bullet, which will be lower than the
bow shock speed.  It is also possible that the bullet responsible for the X-ray knot will be
moving into a medium which itself has been set into motion by the 
preceding bullet, leading to lower shock speeds.

If, however, the jet is a continuous flow, then the optical shape of the X-ray knot
and its coincidence with the X-ray emission leads to another
interesting possibility.  The narrowing of the flow at the knot in Hen\,3-1475 is
not seen in other jet-producing environments and has been noted as
a potential example of a ``conical converging flow" in which
material is inertially focused by the walls of an axisymmetric
cavity to converge at a point where shocks redirect the
material into a jet \citep{canto88,frank96}. Recent
laboratory experiments have shown such flows to capable of
producing remarkably stable jets \citep{lebedev02}. While the
S-shaped symmetry evident in He3-1475 makes pure hydrodynamic
collimation unlikely, the presence of toroidal magnetic fields may produce
conically converging flows in a similar manner which would also
show X-rays at the convergence point.  
In fact, Hen\,3-1475 
is the first case of unambiguously shock-excited X-ray emitting gas to 
be detected by CXO that is unresolved (e.g., all of 
the PNs detected by CXO show resolved nebular X-ray emission).
The various jet models for Hen\,3-1475 described above need to be 
tested using detailed (magneto)hydrodynamic simulations 
to quantitatively fit the observed X-ray and optical (line) fluxes and 
spectra.

We are thankful for partial financial support for this 
work provided by NASA through its 
Long Term Space Astrophysics program (grant no. 399-30-61-00-00) for R.S. and M.M., as well as through 
Chandra Awards GO2-3029X (for R.S.) and GO2-3009X (for J.H.K./RIT) issued by 
the Chandra X-ray Observatory Center, 
which is operated by the Smithsonian Astrophysical Observatory 
on behalf of NASA under contract NAS8-39073.

\newpage 

\newpage

\end{document}